\documentclass{aa} 
\usepackage{graphicx}
\usepackage{txfonts}
\def\ltapprox{\raise 2pt \hbox {$<$} \kern-1.1em \lower 5pt \hbox {$\approx$}}
\def\ltsim{\; \raise0.3ex\hbox{$<$\kern-0.75em \raise-1.1ex\hbox{$\sim$}}\; }
\def\gtsim{\; \raise0.3ex\hbox{$>$\kern-0.75em \raise-1.1ex\hbox{$\sim$}}\; }

\def\ie{{\it i.e.,~}}

\def\eg{{\it e.g.,~}}

\begin{document}
   \title{On the radio -- X-ray luminosity correlation of radio halos at low radio frequency}
   \subtitle{Application of the turbulent re-acceleration model}

   \author{R. Cassano\inst{1}\fnmsep\thanks{\email{rcassano@ira.inaf.it}}}
%	   G. Brunetti\inst{1}}
%	   G. Brunetti\inst{1}}
\authorrunning{R. Cassano}

   \offprints{R.Cassano}

   \institute{INAF - Istituto di Radioastronomia, via P. Gobetti 101,I-40129 Bologna, Italy
%	\and
%	INAF - Osservatorio Astronomico di Bologna,
%via Ranzani 1, I-40127 Bologna, Italy
}
%   \and
%   Dipartimento di Astronomia, Universit\`a di Bologna, via Ranzani 1, I-40127 Bologna, Italy

%\date{Received...; accepted...}

\abstract
% context heading (optional)
{} 
% aims heading (mandatory)
{In this paper we show expectations on the radio--X-ray luminosity correlation of radio halos at 120 MHz. According to the ``turbulent re-acceleration scenario'', low frequency observations are expected to detect a new population of radio halos that, due to their ultra-steep spectra, are missed by present observations at $\sim$ GHz frequencies. These radio halos should also be less luminous than presently observed halos hosted in clusters with the same X-ray luminosity.} 
{Making use of Monte Carlo procedures, we show that the presence of these ultra-steep spectrum halos at 120 MHz causes a steepening and a broadening of the correlation between the synchrotron power and the cluster X-ray luminosity with respect to that observed at 1.4 GHz.} 
{We investigate the role of future low frequency radio surveys, and find that the upcoming LOFAR surveys will be able to test these expectations.} 
{}

%We show that with increasing of the sensitivity of LOFAR observations a large number of ultra-steep spectrum halos will be find and this will make even steeper the correlation.
         

\maketitle

\section{Introduction}

Radio halos are diffuse synchrotron sources from the intra--cluster medium (ICM) extended on mega-parsec scale (\eg Feretti 2005; Ferrari et al. 2008). They provide the most important evidence of non-thermal components (relativistic particles and magnetic fields) mixed with the hot ICM.
Galaxy clusters hosting radio halos are always characterized by a non-relaxed dynamical status suggestive of recent or ongoing merger events (\eg Buote 2001; Schuecker et al 2001; Govoni et al. 2004; Venturi et al. 2008; Giacintucci et al. 2009). Furhermore, the halo radio power at 1.4 GHz increases with the cluster X-ray luminosity, mass and temperature (\eg Liang et al. 2000; En\ss lin \& R\"ottgering 2002; Bacchi et al. 2003; Clarke 2005; Dolag et al. 2005; Cassano et al. 2006, 2007; Brunetti et al. 2009; Rudnick \& Lemmerman 2009; Giovannini et al. 2009). These correlations and the radio halo-merger connection suggest that gravity provides the reservoir of energy to generate the non-thermal components (\eg Kempner \& Sarazin 2001). Cluster mergers drive shocks and turbulence in the ICM that may amplify the magnetic fields (\eg Carilli \& Taylor 2002; Dolag et al. 2002; Br\"uggen et al. 2005; Subramanian et al. 2006; Ryu et al. 2008) and accelerate high energy particles (\eg Fujita et al. 2003; Hoeft \& Br\"uggen 2007; Brunetti \& Lazarian 2007; Pfrommer et al. 2008; Vazza et al. 2009).

Two main scenarios have been proposed to explain the origin of relativistic particles in radio halos, namely {\it i)} the {\it turbulent re-acceleration} model,
whereby relativistic electrons are re-energized {\it in situ} due to the interaction with MHD turbulence generated in the ICM during cluster mergers (\eg Brunetti et al. 2001; Petrosian et al. 2001), and {\it ii)} the {\it secondary electron} models, whereby the
relativistic electrons are secondary products of the collisions between cosmic rays and thermal protons in the ICM (\eg Dennison 1980; Blasi \& Colafrancesco 1999, Pfrommer \& En\ss lin 2004).
  
Observations provide support to the idea that turbulence may play a role in the particle re-acceleration process (\eg Brunetti et al. 2008; Ferrari et al. 2008; Cassano 2009; Giovannini et al. 2009), in which case, the population of radio halos is predicted to be a mixture of sources with different spectral properties, with halos having steeper spectra being more common (Cassano et al. 2006a, hereafter C06; Cassano et al. 2009, hereafter C09). In this respect, since very steep spectrum halos should glow up at low radio frequency, upcoming observations with the Low Frequency Array (LOFAR) and the Long Wavelength Array (LWA) will be crucial.

In this paper we discuss how such a predicted population is expected to affect the properties of the radio--X-ray luminosity correlation at low radio frequency, and investigate the potential of LOFAR surveys.
A $\Lambda$CDM cosmology ($H_{o}=70\,\rm km\,\rm s^{-1}\,\rm Mpc^{-1}$, $\Omega_{m}=0.3$,
$\Omega_{\Lambda}=0.7$) is adopted.

\section{The population of ultra-steep spectrum radio halos}
\label{Sect:model}

The formation and evolution of radio halos according to the {\it turbulent re-acceleration scenario} have been investigated by means of Monte Carlo based procedures (Cassano \& Brunetti 2005; C06; C09). These procedures allow us to account for the main ingredients in the model, \ie the rate of cluster-cluster mergers in the Universe, their mass ratio, and the fraction of energy dissipated during mergers that is channelled into MHD turbulence and acceleration of relativistic particles.
We simulate the formation history of $\sim 1000$ galaxy clusters with present day masses in the range
$\sim[0.2-6]\times 10^{15}\,M_{\odot}$.

Turbulence acceleration is a rather inefficient process in the ICM and electrons can be accelerated up to energy of several GeV, since at higher energy the radiation losses dominate (\eg Brunetti \& Lazarian 2007). This implies the presence of a gradual spectral steepening at high frequencies in the synchrotron spectrum of radio halos. The presence of this steepening makes it difficult to detect radio halos at frequencies larger than the frequency $\nu_s$ at which the steepening becomes severe. 

Following C06 and C09 we use homogeneous models that assume: {\it i)} an average value of the magnetic field strength in the radio halo volume that scales with the cluster mass as $B=B_{<M>} (M_v/<M>)^b$ \footnote{$B_{<M>}$ is the value of the magnetic field averaged in a region of radius $=500\,h_{50}^{-1}$ kpc in a cluster with viral mass $<M>=1.6\times10^{15}\,M_{\odot}$.}, {\it ii)} that a fraction, $\eta_t$, of the $P dV$ work, done by subclusters crossing the main clusters during mergers goes into {\it magneto-acoustic} turbulence. The frequency $\nu_s$, defined as the frequency at which the spectral slope of the halos becomes $\alpha\geq1.9$ ($P(\nu)\propto \nu^{-\alpha}$), depends on the acceleration efficiency in the ICM, $\chi$, on the magnetic field in the ICM, $B$, and on the energy density of the cosmic microwave background radiation (CMB) as: $\nu_s\propto (B\,\chi^2)/(B^2+B_{cmb}^2)^2$. The frequency $\nu_s$ is a more practical re-definition of the synchrotron break frequency, $\nu_b$, and, in the case of homogeneous models is $\nu_s\sim 7\,\nu_b$ (C09). According to C09, in the case of a single major merger between a cluster of mass $M_v$ and a subcluster with mass $\Delta M$, $\nu_s$ is 
 
\begin{equation}  
\nu_s\propto \frac{B\,(k_BT)^{-1}}{(B^2+B_{cmb}^2)^2}\Big(\frac{M_v+\Delta M}{R_v}\Big)^3\,,
\label{Eq:nub}
\end{equation}

\noindent where $B_{cmb}=3.2 (1+z)^2 \mu G$ is the equivalent magnetic field strength of the CMB, and $R_v$ is the cluster virial radius. It is expected that mergers may generate halos with larger $\nu_s$ in more massive clusters, and that halos in clusters with the same mass $M_v$ (and magnetic field) and redshift could have different $\nu_s$ depending on the properties of the merger event responsible for their generation. 
Halos with $\nu_s \geq$ 1.4 GHz must be generated in connection with the most energetic merger-events in the Universe since only these mergers may allow for the efficient acceleration that is necessary to have relativistic electrons emitting at these frequencies. 
Present surveys carried out at $\nu_o \sim$ 1 GHz detect radio halos only in the most massive and merging clusters (\eg Buote 2001, Venturi et al. 2008). On the other hand, radio halos with smaller values of $\nu_s$ must be more common, since they can be generated in connection with less energetic phenomena, \eg major mergers between less massive systems or minor mergers in massive systems, that are more frequent in the Universe. This has been addressed quantitatively by means of Monte Carlo calculations that allow us to derive the fraction of clusters with radio halos with different $\nu_s$ as a function of the cluster mass and redshift. The expected population of radio halos is indeed constituted by a mixture of halos with different spectra, with steep spectrum halos being more common in the Universe; $\sim 200$ radio halos with $120<\nu_s<600$ MHz are expected in future LOFAR surveys at 120 MHz (C09).

\begin{figure}
%\begin{center}
\centerline{
\includegraphics[width=0.4\textwidth]{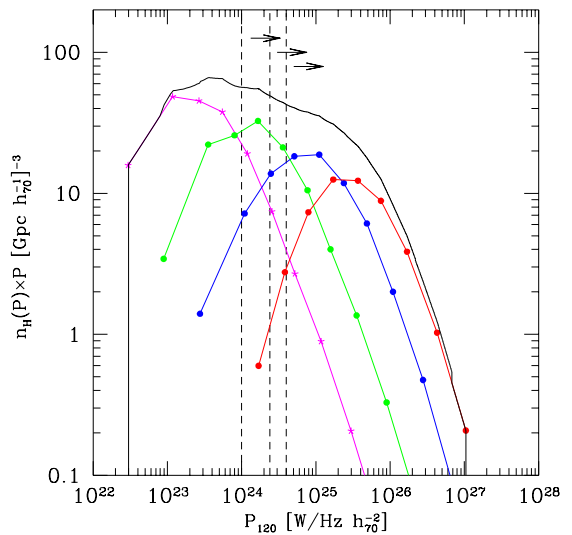}}
\caption[]{RHLF at 120 MHz (black lines) and in the redshift interval $z=0-0.1$. The differential contributions from halos with $120<\nu_s<240$ MHz (magenta line, star symbols), $240<\nu_s<600$ MHz (green line), $600<\nu_s<1400$ MHz (blue line) and $\nu_s>1400$ MHz (red line) are also shown. The dashed lines correspond to the minimum detectable halo power for $\xi\cdot F=0.25, 0.6$ and $1$ mJy/beam (from left to right, see text).}
\label{Fig.RHLF}
%\end{center}
\end{figure}

According to this model the monochromatic luminosity of radio halos at a given frequency, $\nu_0<\nu_s$, increases with increasing $\nu_s$. 
For a fixed cluster mass (or X-ray luminosity), the relation between the radio luminosity at $\nu_0$ of halos with $\nu_s=\nu_{s,1}$ and $\nu_{s,2}$, $P_{\nu_{s,1}}(\nu_0,L_X)$ and $P_{\nu_{s,2}}(\nu_0,L_X)$ respectively, is (C09)

\begin{equation}
P_{\nu_{s,1}}(\nu_o, L_X)=
P_{\nu_{s,2}}(\nu_o, L_X) \Big(\frac{\nu_{s,1}}{\nu_{s,2}}\Big)^{\alpha}\,.
\label{Eq:Pnu_P1p4}
\end{equation}

%For a fixed cluster mass (or X-ray luminosity), the relation between the radio luminosity at $\nu_0$ of halos with $\nu_s\ge 1.4$ GHz, $P_{1.4}(\nu_o)$, and that of halos with a given value of $\nu_s$, $P_{\nu_s}(\nu_0)$, is (C09)

%\begin{equation}
%P_{\nu_s}(\nu_o, L_X)=
%P_{1.4}(\nu_o, L_X) \Big(\frac{\nu_s}{1400\,\mathrm{MHz}}\Big)^{\alpha}\,.
%\label{Eq:Pnu_P1p4}
%\end{equation}
\noindent where $\alpha\approx 1.3$ (\eg Ferrari et al. 2008) is the radio spectral index. A correlation between the radio power $P_R$ of halos and the mass (and X-ray luminosity) of the hosting clusters is expected (C06). In the simplest case that halos are generated by a single major merger this is

\begin{equation}  
P_{R}\propto \frac{M_v^{2-\Gamma}\,B^2}{(B^2+B_{cmb}^2)^2}\,,
\label{Eq:PM}
\end{equation}

\noindent where the parameter $\Gamma$ is defined by $T\propto M^{\Gamma}$ ($\Gamma\simeq 2/3$ in the virial scaling). Eq.~\ref{Eq:PM} implies that more massive clusters host more luminous radio halos. By considering halos with $\nu_s\geq1.4$ GHz, C06 showed that the slope of this scaling is consistent with that of the observed $P(1.4)-M_v$ correlation, provided that the model parameters $(B_{<M>}, b , \eta_t)$ lie within a fairly constrained range of values (see Fig.~7 in C06). We refer the reader to Sects. 3.3 and 4.1 of C06 for a more detailed discussion on model parameters and on their constraints. 

Following C09, we adopt a reference set of parameters\footnote{We note that for this particular configuration of parameters even in the case of re-acceleration phase the energy of magnetic field is always dominant with respect to that of relativistic electrons.}: $B_{<M>}=1.9\, \mu$G, $b=1.5$, $\eta_t=0.2$, that falls in that range and sets $\alpha_M=3.3$, with $P(1.4)\propto M_v^{\alpha_M}$. This implies $P(1.4)\propto L_X^{2.25}$ assuming the $L_X-M_v$ correlation for galaxy clusters as derived in C06.  
Since the bulk of radio halos in our calculations is found to be associated with clusters of mass $\sim[1-2]\times 10^{15}\,M_{\odot}$ the adopted values of $B$ and $b$ imply typical average magnetic fields $\sim 1-3$\,$\mu$G. These values of $B$ are similar to those derived from rotation measurements (\eg Govoni \& Feretti 2004; Bonafede et al. 2010) and equipartition assumption (En\ss lin et al. 1998; Govoni et al. 2001).

\begin{figure*}
\begin{center}
\includegraphics[width=0.33\textwidth]{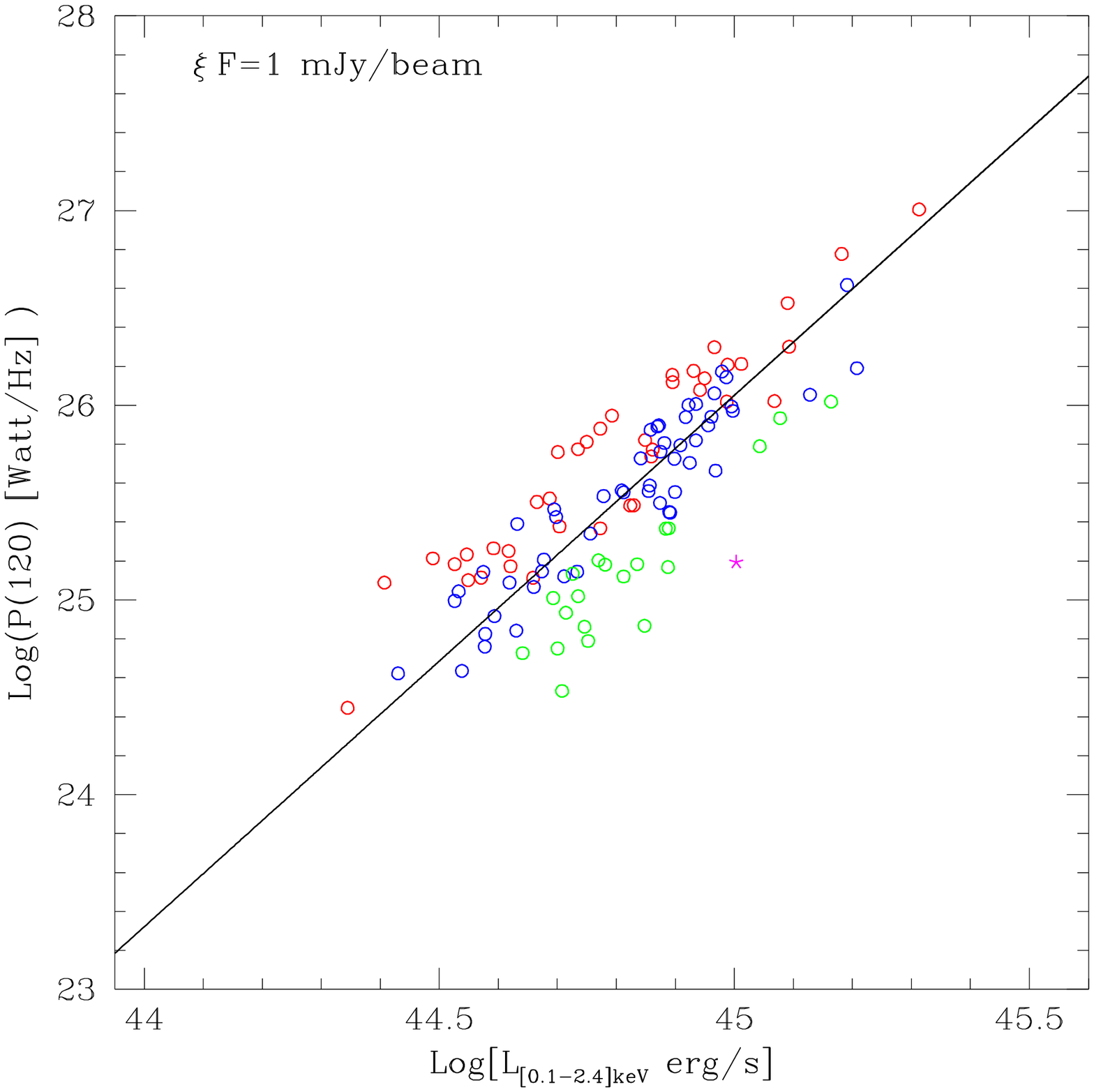}
\includegraphics[width=0.33\textwidth]{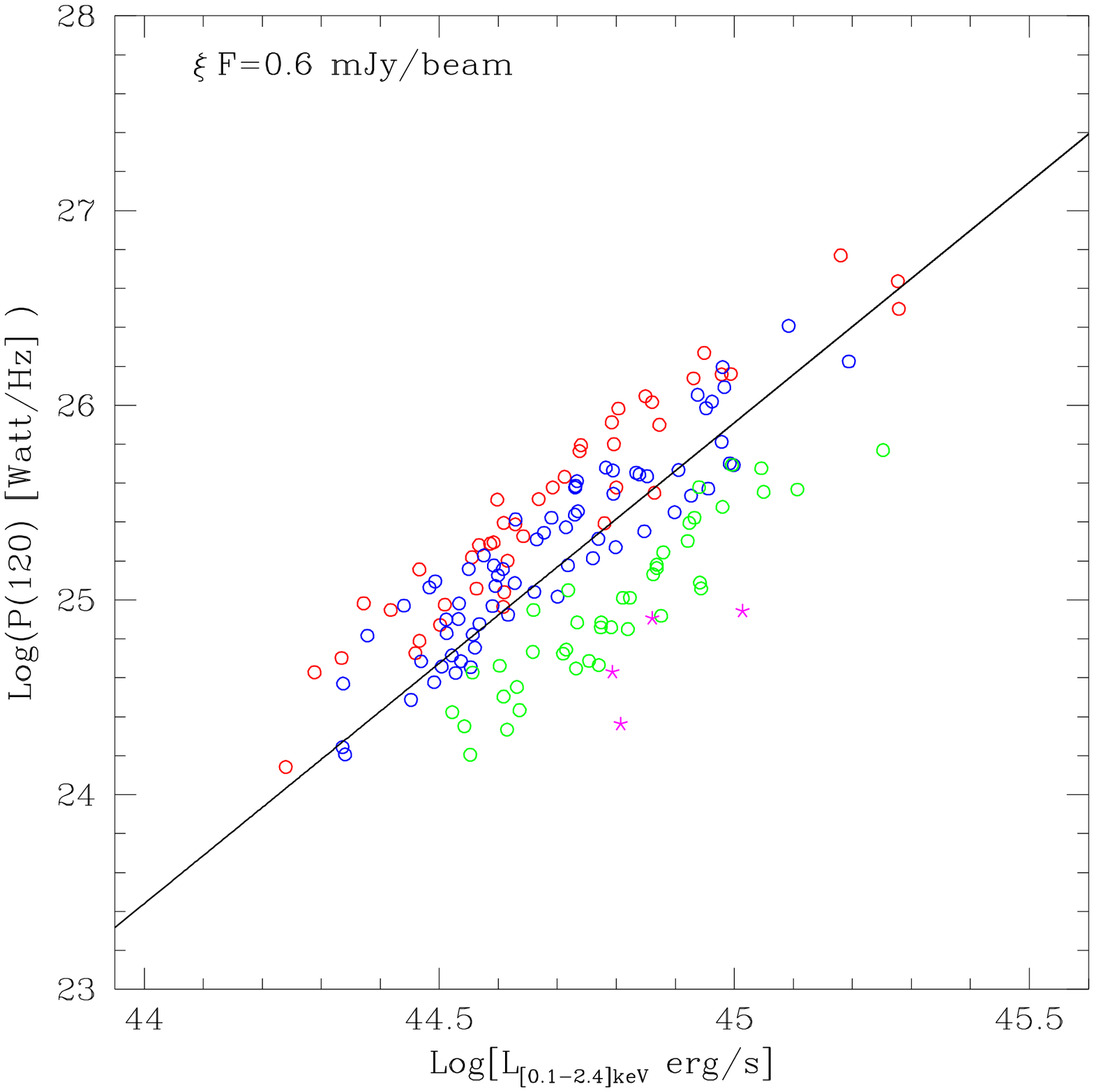}
\includegraphics[width=0.33\textwidth]{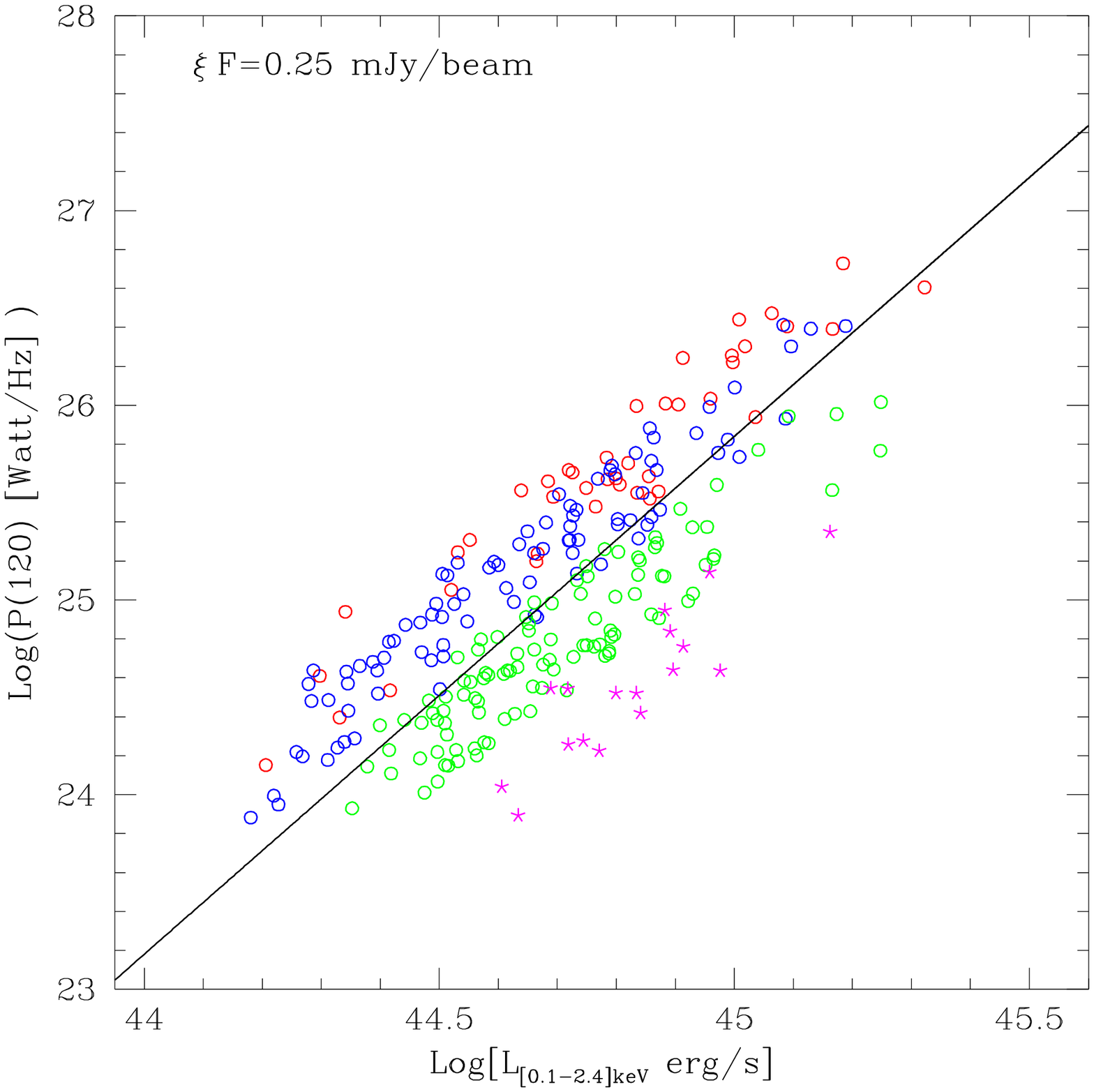}
\caption[]{Expected distribution of radio halos in the $P(120)-L_X$ diagram (colored open dots, the color code is the same as Fig.~\ref{Fig.RHLF}) together with the position of radio halos observed at 1.4 GHz (black filled dots). The correlation $P(1.4)-L_X$ extrapolated at 120 MHz (dashed lines) and the best fit of the $P(120)-L_X$ correlation (solid lines) are also shown. From left to right $\xi\,F=1$, $0.6$ and $0.25$ mJy/beam.}
\label{Fig.Lr_Lx}
\end{center}
\end{figure*}

The observed $P(1.4)-L_X$ correlation shows an intrinsic scatter across the radio luminosity $\delta P/P\simeq \pm 2$ (\eg Brunetti et al. 2009). In principle, in our model a scatter in the $P(1.4)-L_X$ correlation is expected due to the different monochromatic radio luminosity of halos with different $\nu_s$ (Eq.~\ref{Eq:Pnu_P1p4}). Our calculations show that the fraction of clusters hosting radio halos with $\nu_s\ge 3500$ MHz is about a few percent, thus we would expect $\delta P/P=\pm 1/2\times(3500/1400)^{1.3}\sim 1.7$ for halos with $\nu_s\ge 1.4$ GHz, which is in line with the observed scatter. However, we stress that there are other possible sources of scatter which are difficult to take into account in homogeneous models. These are due to \eg differences in the 
cosmic rays and magnetic field content in clusters with the same mass. 
 
Once we anchor the luminosity of halos with $\nu_s>1.4$ GHz, $P_{1.4}(1.4,L_X)$, to the observed $P(1.4)-L_X$ correlation, Monte Carlo calculations carried out by considering an observing frequency $\nu_0$ and Eq.~\ref{Eq:Pnu_P1p4} allow us to derive the expected radio halo luminosity functions (RHLF; see \eg C06 \& C09 for details). As an example, Fig.~\ref{Fig.RHLF} shows the total RHLF obtained from Monte Carlo calculations at $\nu_0=120$ MHz (black line) and $z=0-0.1$, together with the differential contributions to the RHLF from halos with different $\nu_s$ (see figure caption). As expected, radio halos with smaller $\nu_s$ mainly contribute to the low-power end of the total RHLF, and the peaks of the RHLF of different populations move towards low radio powers with decreasing $\nu_s$. This implies that, depending on their sensitivity, surveys at low radio frequency will unveil new populations of halos.

\section{Monte Carlo distributions of radio halos in the $P(120)-L_X$ plane}
\label{Sect:corr}

The aim of this section is to investigate how the presence of the new population of ultra-steep spectrum halos, predicted in deep low frequency radio surveys, may affect the radio -- X-ray luminosity correlation of halos at low radio frequency. LOFAR will carry out surveys between 15 and 210 MHz in the Northern hemisphere with unprecedented sensitivity and spatial resolution. Since LOFAR is expected to carry out the deepest large area radio surveys at $\nu_o$= 120 MHz (\eg R\"ottgering et al. 2006), in this paper we focus on the $P(120)-L_X$ correlation.

\begin{figure*}
\begin{center}
\includegraphics[width=0.36\textwidth]{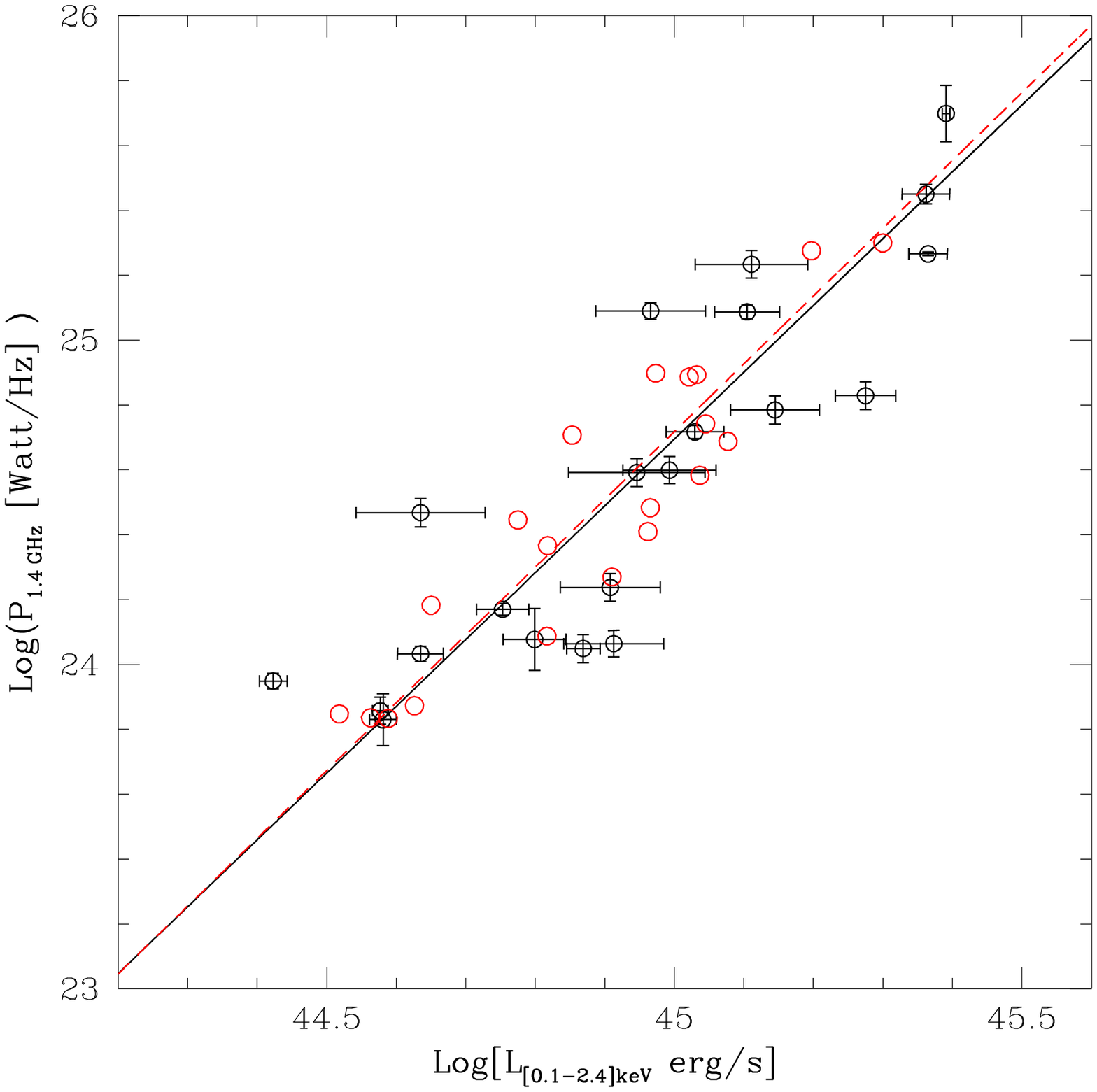}
\includegraphics[width=0.36\textwidth]{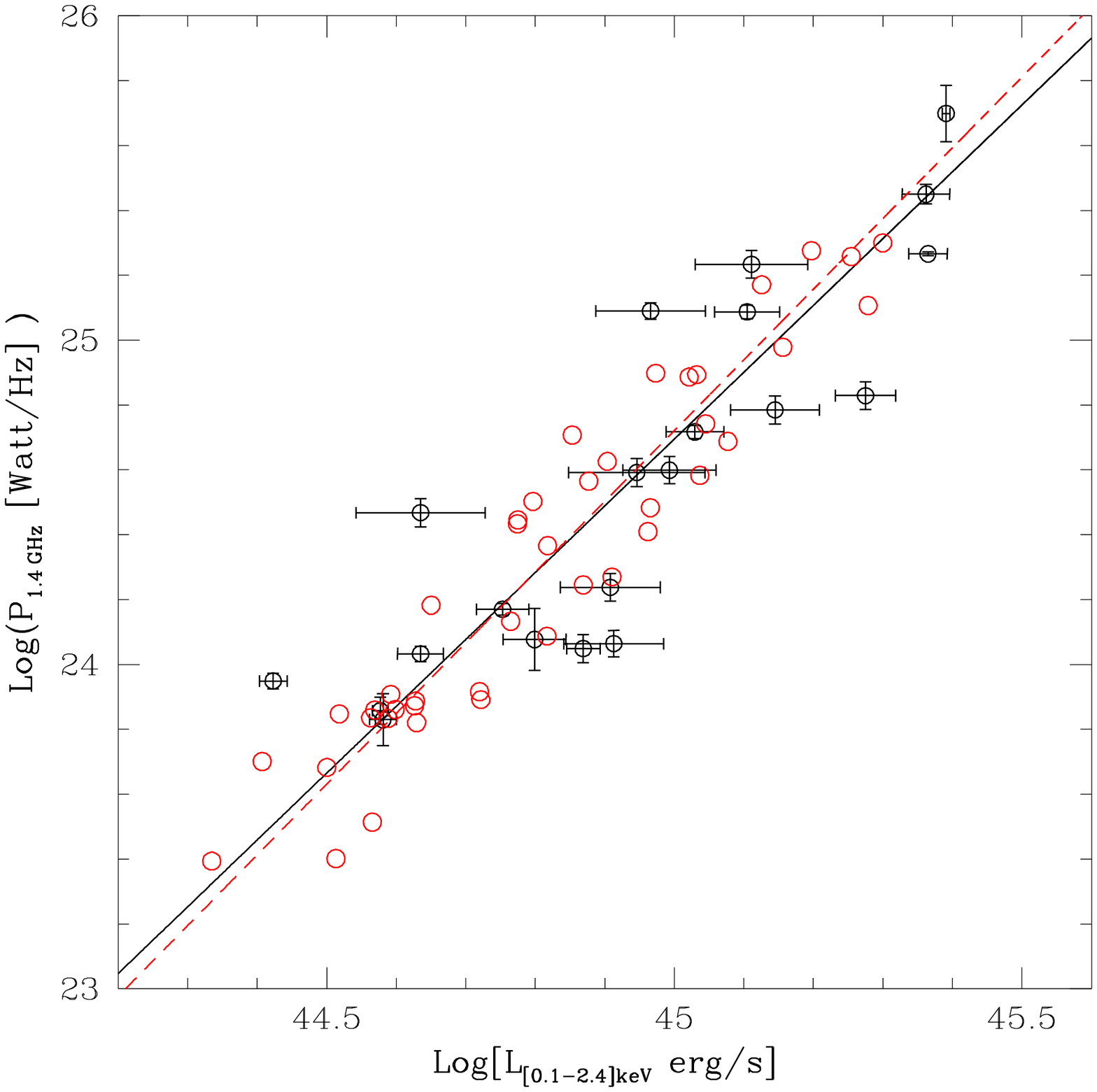}
\caption[]{Radio -- X-ray luminosity correlation of giant radio halos at 1.4 GHz. Observed halos at 1.4 GHz (black points) and ``simulated halos'' with $\nu_s>1.4$ GHz (red points) are plotted together with their best-fit relations (solid black line and dashed red line, respectively). {\it Left Panel}: the observed halos are compared with those expected by the model by considering X-ray and radio sensitivities, and sky coverage of GMRT+NVSS samples (see text). {\it Right Panel}: present observed halos are compared with those, with $\nu_s>1.4$ GHz, that are expected to be detected by LOFAR surveys assuming $\xi\,F\simeq 0.25$ mJy/beam.}  
\label{Fig.LrLx_1p4}
\end{center}
\end{figure*}

The most crucial point in this respect is the estimate of the minimum diffuse flux of a radio halo detectable by these surveys as a function of redshift. It is well known that the brightness profiles of radio halos smoothly decrease with distance from the cluster center (\eg Govoni et al. 2001; Murgia et al. 2009) implying that the outermost region of the halos will be difficult to detect in radio surveys. 
Brunetti et al. (2007) found that the typical profiles of radio halos are such that about $58$\% of their flux is contained within half radius ($R_H$). Following C09, in order to have a good sensitivity to diffuse emission, we assume a beam of $25\times25$ arcsec and estimate the minimum flux of a Mpc-sized halo detectable in a LOFAR survey, $f_{H}$, by requiring that the mean brightness within half radius of a halo, $B_{<0.5\,R_H}$, is $\xi$ times the $rms$ ($F$) of the survey, \ie

\begin{eqnarray}
B_{<0.5\,R_H}\simeq \frac{0.58 f_H}{\pi/4\, \theta_{H}^2}=\xi \, F \,\,\Rightarrow \nonumber\\
%\Rightarrow
%\end{equation}
%\begin{equation}
f_{H}(z)\simeq 2\times 10^{-3}\Big[\frac{\xi \, F}{\mathrm{mJy/b}}\Big]\,
\Big[\frac{\theta_{H}^2(z)}{\mathrm{arcsec^2}}\Big] \,\, \mathrm{mJy}\,,
\label{fmin}
\end{eqnarray}

\noindent
where $\theta_{H}(z)$ is the angular size of radio halos, in arcseconds, at a given redshift. 
In the case of $\xi\simeq1$ this approach would guaranty the detection, at several $\sigma$, of the central brightest region of halos, thus leading to the identification of candidate radio halos in the survey.

This simple approach has been tested in Cassano et al. (2008) by injecting ``fake'' radio halos in the (u,v) plane of NVSS and GMRT observations. It has been shown that radio halos become visible in the images as soon as their flux approaches that obtained by Eq.~\ref{fmin} with $\xi=1$ and $\xi=2$ for the NVSS and GMRT observations, respectively (see also Brunetti et al. 2007; Venturi et al. 2008).

The LOFAR Large Area Survey is expected to reach an $rms\sim0.1$ mJy/beam at 120 MHz with a beam  $\sim5\times$5 arcsec. These resolution and sensitivity are obtained by assuming a ``uniform weighting'', thus we may assume that the same sensitivity can be reached with a larger beam, obtained for example by \eg ``natural weighting''.
%In C09 a tapered beam of 25$\times$25 arcsec is assumed with the same rms sensitivity. This assumption is reasonable due to the large number of short baselines of LOFAR.
However, the survey sensitivity may be limited by the rms confusion level. This is given by (\eg Condon 1987; Kronberg et al. 2007):

\begin{eqnarray}
\sigma_{conf}\simeq 0.13 \, \times \Big[\frac{\theta_1 \times \theta_2}{25\times 25 \,\mathrm{arcsec}^2}\Big] \times \Big(\frac{\nu}{120\,\mathrm{MHz}}\Big)^{-0.7} \, {\mathrm m \mathrm Jy} 
\label{sigma_conf}
\end{eqnarray}

\noindent where $\theta_{1,2}$ is the beam size in arcsec, and $\nu$ is the frequency in MHz.
%, however it can be shown that the rms confusion level (Condon 1987) may limit the survey sensitivity with that tapered beam at rms $\geq 0.1$ mJy/beam.

Of course all the issues discussed above will be clarified during the commissioning phase of LOFAR. Thus we decided to present calculations in several cases, specifically  $\xi\cdot F\,=0.25$, $0.6$ and $1$ mJy/beam to cover a range of possible LOFAR sensitivities \footnote{Note that our choice of $\xi\cdot F$ 
is thought to mimic different possible configurations, \eg $\xi\cdot F=0.6$ mJy/beam can be $\xi=3$ (3$\sigma$ detection of the average halo brightness in half $R_H$) and 
F=$0.2$ mJy/beam, or $\xi=1$ and F=$0.6$ mJy/beam}. 
Vertical dashed lines in Fig.~\ref{Fig.RHLF} show the minimum power of a halo at $z\sim 0.05$ detectable by LOFAR surveys assuming $\xi\cdot F=0.25, 0.6$ and $1$ mJy/beam. The important point is that with increasing survey sensitivity new populations of radio halos are expected to be unveiled, with the detectable number of ultra steep spectrum halos increasing in deeper surveys.

LOFAR observations will allow to study the distribution of radio halos in the radio--X-ray luminosity diagram at low radio frequencies, so far an unexplored issue. The vast majority of ultra-steep spectrum halos visible at low frequencies are expected to be associated with galaxy clusters of intermediate X-ray luminosity, $L_X\sim 3-6\cdot 10^{44}$ erg/s, and should be less luminous than radio halos that are presently observed at GHz frequencies. This should affect the radio--X-ray luminosity correlation of halos at low frequencies, that is expected to be steeper and with larger scatter than that at 1.4 GHz.

To address this issue quantitatively we assume $\nu_0=120$ MHz, and following C06 and C09 we use Monte Carlo procedures based on the extended Press \& Schechter (1974; Lacey \& Cole 1993) formalism to obtain {\it i)} the population of galaxy clusters, with their mass (and X-ray luminosity), in the redshift interval $z=0-0.5$, and {\it ii)} the population of radio halos, with their $\nu_s$, associated with these clusters. We use homogeneous models and the set of model parameters given in the previous section. From these simulations we extract the population of radio halos that can be detected by observations at $\nu_0=120$ MHz according to their radio luminosity and $f_{min}(z)$. 

In particular, the luminosity at 120 MHz of radio halos with $\nu_s\geq 1.4$ GHz, in clusters with X-ray luminosity $L_X$, is obtained from the $P(1.4)-L_X$ correlation, assuming a spectral index $\alpha=1.3$ and allowing for a random scatter $\delta P/P=\pm 2$ (see discussion in Sect.2).
The luminosity at 120 MHz of radio halos with a given $\nu_s$ is obtained according to Eq.~\ref{Eq:Pnu_P1p4}. In particular, we calculated halo statistics by assuming the following frequency ranges: $\nu_s=120-240$ MHz, $240-600$ MHz, $600-1400$ MHz.
Eq.~\ref{Eq:Pnu_P1p4} also implies that halos with $\nu_{1}\leq\nu_s<\nu_{2}$ should have radio luminosities at 120 MHz which may scatter by a factor $(\nu_{2}/\nu_{1})^{\alpha}$ that implies $\delta P/P\simeq\pm 1.3-1.7$ for the frequency bins we are considering.
%Since we are considering similar range of $\nu_{s,2}/\nu_{s,1}$, this implies a scatter in the 120 MHz radio luminosity in the range $\delta P/P\simeq\pm 1.3-1.7$ (Eq.~\ref{Eq:Pnu_P1p4}), thus for semplicity we assume a scatter $\delta P/P\simeq\pm 2$.
Finally, we assume the LOFAR sky coverage (the Northern hemisphere, $\delta\geq0$, and high Galactic latitudes, $|b|\geq 20$) and $f_{min}(z)$ from Eq.~\ref{fmin}.

\begin{figure}
%\begin{center}
\centerline{\includegraphics[width=0.36\textwidth]{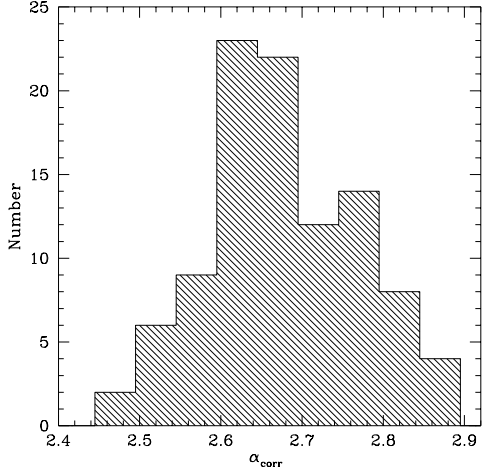}}
\caption[]{Spectral slopes of the $P(120)-L_X$ correlation obtained after 100 Monte Carlo extractions of the radio halo distribution in the $P(120)-L_X$ diagram, assuming $\xi\,F=0.25$ mJy/beam.}
\label{Fig.alpha}
%\end{center}
\end{figure}

The resulting theoretical distribution of radio halos in the $P(120)-L_X$ diagram is shown in Fig.~\ref{Fig.Lr_Lx}, assuming $\xi\cdot F=1, 0.6$ and $0.25$ mJy/beam (colored open dots; from left to right).
%, together with the distribution of giant radio halos presently observed at 1.4 GHz, whose radio luminosity at 120 MHz has been extrapolated by using $\alpha=1.3$ (black filled dots). 
Different colored dots indicate halos with different values of $\nu_s$ (the same color code used in Fig.~\ref{Fig.RHLF}).
Halos with different $\nu_s$ fill different regions, with radio halos with smaller $\nu_s$ typically located in regions of lower radio luminosities. 
%As expected, radio halos with $\nu_s\geq1.4$ GHz (red dots) follow a trend consistent with presently observed halos (black points), their larger number reflects the large sensitivity of LOFAR surveys with respect to present surveys. 
The number of halos with smaller $\nu_s$ increases with increasing survey sensitivity. In the case of high sensitivity surveys these halos dominate the population and their presence affects the overall shape of the correlation. The correlation at 120 MHz is predicted more scattered and steeper than that observed at 1.4 GHz. A quantitative estimates of the steepening  can be obtained by repeating many times the Monte Carlo procedure described above and by fitting the obtained halo distributions in the $P(120)-L_X$ diagram.
Fig.~\ref{Fig.alpha} shows a histogram of the slopes of the correlation obtained after $100$ Monte Carlo runs in the case $\xi\,F\sim 0.25$ mJy/beam.
The mean value of the slope is $\alpha_{corr}\simeq2.68$, while we find $\alpha_{corr}\sim 2.45$ and $2.46$ in the cases $\xi\,F\sim 1$ and $0.6$ mJy/beam, respectively (with 68\% of values typically within $\Delta\alpha\sim0.07$). The values of $\alpha_{corr}$ are significantly larger 
%($\Delta\alpha\sim 0.25-0.65$) 
than that at 1.4 GHz.

\begin{figure}
%\begin{center}
\centerline{\includegraphics[width=0.36\textwidth]{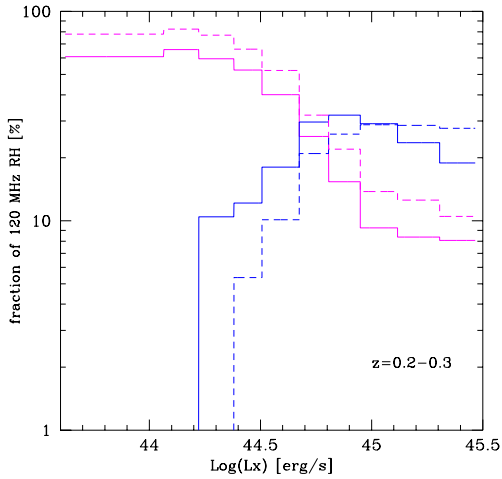}}
\caption[]{Fraction of clusters with radio halos with $120\leq\nu_s<240$ MHz (magenta lines) and $600\leq\nu_s<1400$ MHz (blue lines) as a function of the cluster X-ray luminosity. The calculations are showed for the redshift range $z=0.2-0.3$ and for $B_{<M>}=1.9\, \mu$G, $b=1.5$, $\eta_t=0.2$ (solid lines) and $B_{<M>}=0.2\, \mu$G, $b=0.6$, $\eta_t=0.38$ (dashed lines).}
\label{Fig.USSRH_sub_sup}
%\end{center}
\end{figure}

For completeness, in Fig.~\ref{Fig.LrLx_1p4}, {\it left panel}, we show the theoretical distribution of giant radio halos with $\nu_s>1400$ MHz (red points) in the $P(1400)-L_{X}$ plane together with the observed correlation of halos at 1400 MHz (black points, taken from Brunetti et al. 2009; see Tab.~1 and references therein).
In this case, to compare model expectations and present observations, we follow C09 (Sect.3) and derive the theoretical distribution of radio halos by considering the combination of the NVSS-XBACs (Giovannini et al. 1999) (radio-X-ray) selection criteria and sky coverage (at $z=0.044-0.2$), and the X-ray luminosity range and sky coverage of the GMRT radio halo survey (Venturi et al. 2007, 2008) (at $z=0.2-0.32$)\footnote{The bulk of halos is found in these surveys.}. The observed and theoretical distributions show a good agreement, although data points present a slightly larger scatter than expectations which can be easily interpreted as due to variations of magnetic field in clusters with the same X-ray luminosity (see Sect.~2).
In Fig.~\ref{Fig.LrLx_1p4}, {\it right panel}, the same observed distribution of radio halos at 1.4 GHz (black points) is compared with that of ``simulated'' halos with $\nu_s>1400$ MHz (red points) detectable by a LOFAR survey at 120 MHz with $\xi$ F=0.25 mJy/beam. As expected, radio halos with $\nu_s\geq1.4$ GHz follow a trend consistent with presently observed halos, while their larger number simply reflects the large sensitivity of LOFAR surveys with respect to present surveys.

\subsection{Dependence on model parameters}

The steepening of the correlation is independent of the adopted values of model parameters, at least by considering sets of parameters in the region ($B_{<M>}$, $b$, $\eta_t$) that reproduce both the observed slope of the $P(1.4)-M_v$ correlation ($\alpha_M=2.9\pm0.4$) and the observed fraction of galaxy clusters with radio halos. 
C06 and C09 already discussed the dependence of expectations on model parameters. They showed that the expected number of radio halos decreases only by a factor of $\sim 2-2.5$, from super-linear ($b>1$) to sub-linear ($b<1$) magnetic scaling (see also Fig.~4 in Cassano et al. 2006b). 

For a fixed value of $b$ larger values of $B_{<M>}$ produce $P(1.4)\propto M_v^{\alpha_M}$ correlations with slightly flatter slopes (see Tab.3 in C06). For example, in the case $b=1.5$ the allowed values of $B$ range from $B_{<M>}\simeq1.9\, \mu$G to $\simeq2.8\, \mu$G, and correspondingly it is $\alpha_M=3.3$ and $2.5$, respectively, still consistent with the observed one. 

\begin{figure*}
\begin{center}
\includegraphics[width=0.32\textwidth]{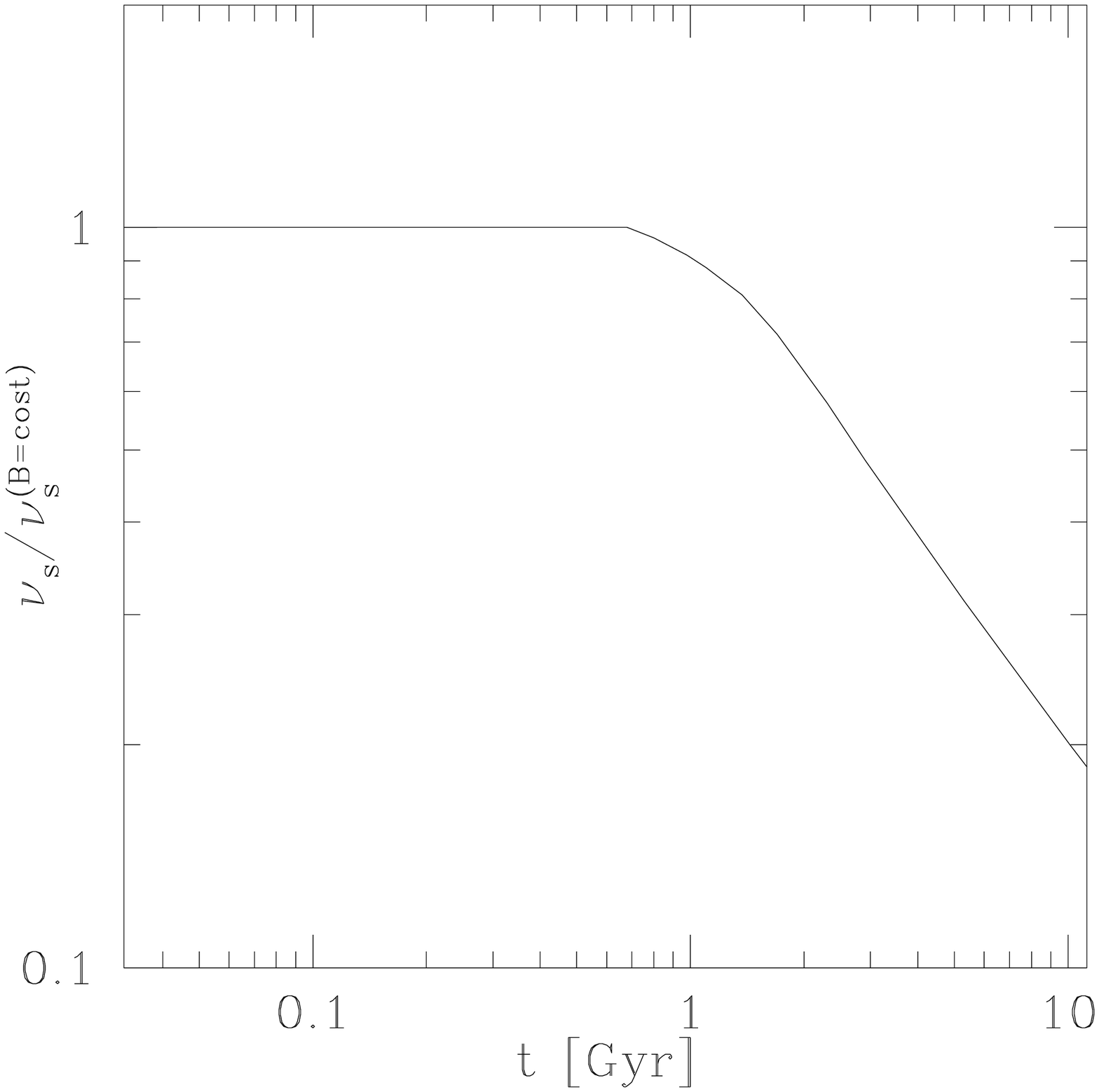}
\includegraphics[width=0.32\textwidth]{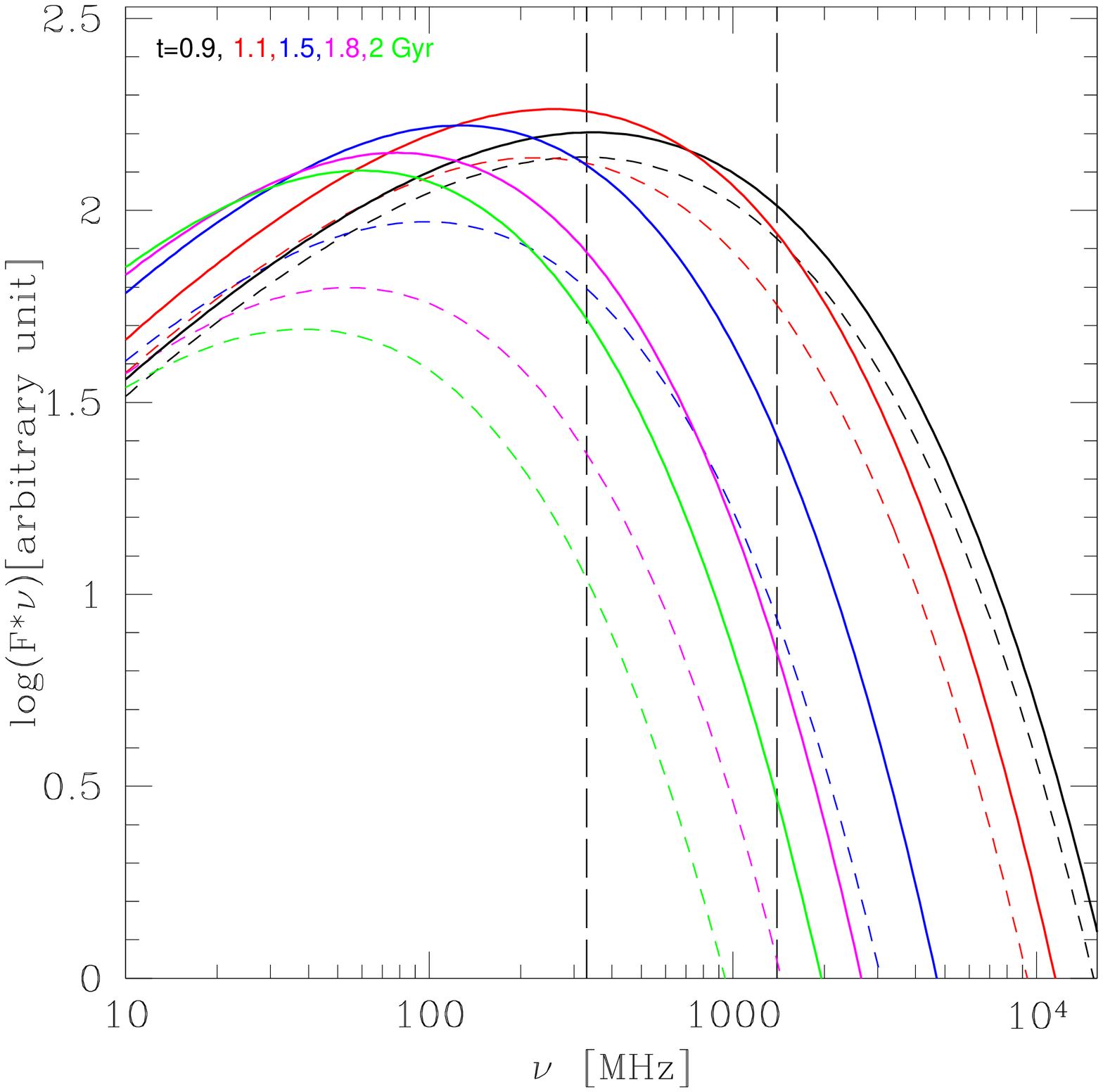}
\includegraphics[width=0.32\textwidth]{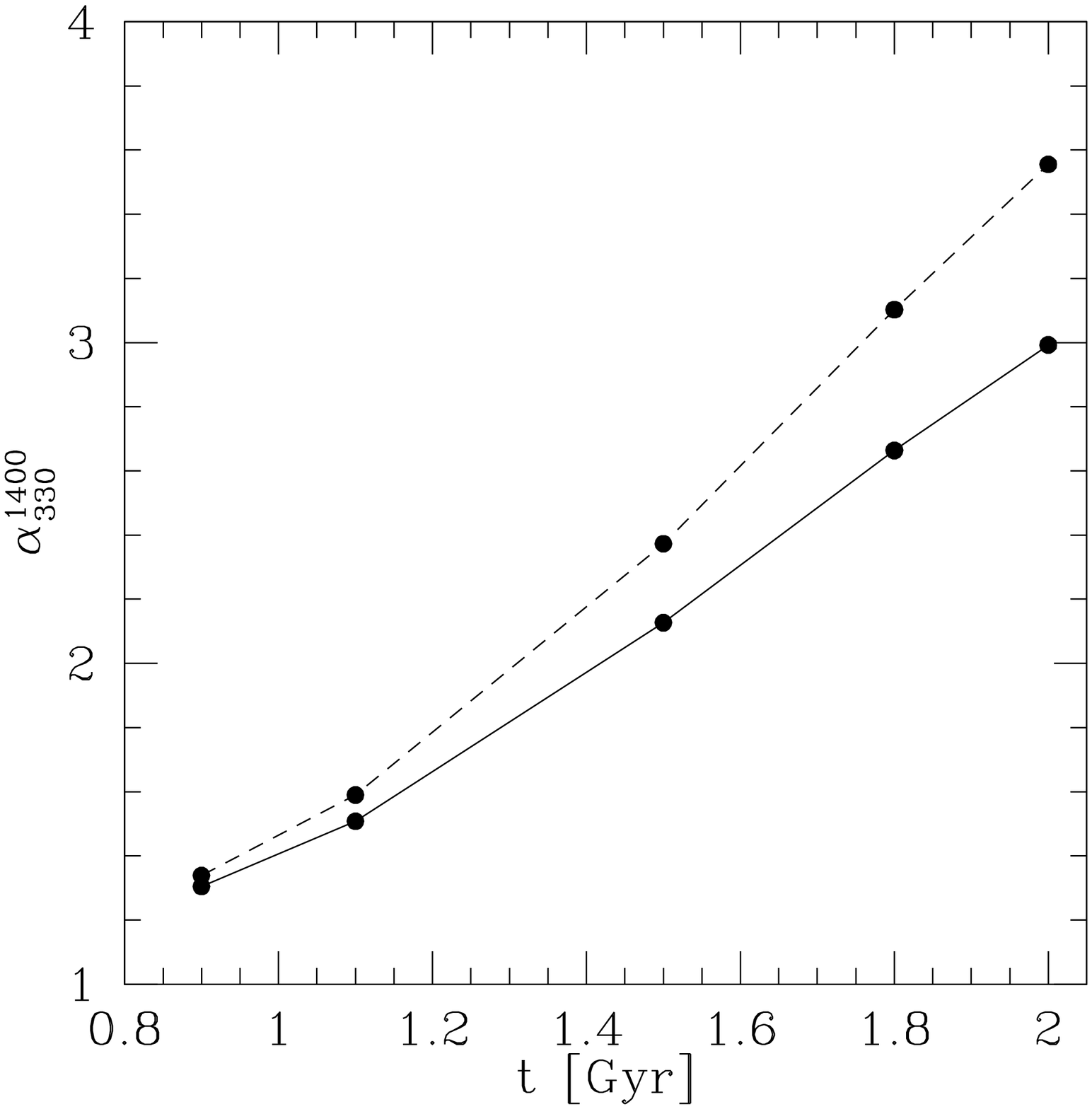}
\caption[]{{\it Left Panel} time-evolution of the ratio between $\nu_s$ computed by assuming an evolving (as in Fig.2 of SSH06) and a constant magnetic field; {\it Central Panel} time-evolution of the synchrotron spectra computed by assuming an evolving (dashed lines) and a constant (solid lines) magnetic field. Spectra are reported for $t=0.9$, $1.1$, $1.5$, $1.8$ and $2$ Gyr (see color code in the figure panel). {\it Right Panel} synchrotron spectral indices between $1.4$ GHz and $330$ MHz of the synchrotron spectra reported above: for a constant magnetic field (solid line) and for an evolving magnetic field (dashed line).}
\label{Fig.evolv_B}
\end{center}
\end{figure*}

The steepening of the correlation is due to the glow up of new radio halos at low frequency, thus another point is whether the fraction of halos with smaller $\nu_s$ (ultra steep spectrum halos) changes from super-linear to sub-linear cases. To investigate this effect, in Fig.~\ref{Fig.USSRH_sub_sup} we report the percentage of radio halos with $120\leq\nu_s<240$ MHz (magenta lines) and $600\leq\nu_s<1400$ MHz (blue lines) as a function of the cluster X-ray luminosity. In Fig.~\ref{Fig.USSRH_sub_sup} we assume two configurations of parameters : the one used in the present paper (solid lines), and a sub-linear one ($B_{<M>}=0.2\, \mu$G, $b=0.6$, $\eta_t=0.38$, dashed lines). In both cases, the vast majority of radio halos hosted in clusters with $L_X\ltsim 3\times10^{44}$ erg/s has $\nu_s\leq 240$ MHz, while halos with $\nu_s\geq 600$ MHz become dominant in more luminous clusters. On the other hand, 
we find that the fraction of halos with $120<\nu_s<240$ MHz is larger in the sub-linear case. This is because it is more difficult to generate radio halos with larger $\nu_s$ in the case of lower magnetic fields (provided that radiative losses are dominated by the Inverse Compton losses due to the CMB photons). We may conclude that in sub-linear cases we expect the following main effects: {\it i)} the $P(120)-L_X$ diagram should be less populated than that in super-linear cases (less halos are expected); {\it ii)} the $P(1.4)-L_X$ correlation is expected to be flatter than that in super-linear cases (see Tab.~3 in C06); {\it iii)} the $P(120)-L_X$ diagram should be even more dominated by ultra-steep spectrum halos, making the steepening of the correlation at lower frequency even stronger.

\section{The effect of an evolving magnetic field}

In this paper we have applied a statistical model based on the turbulence acceleration scenario (discussed and developed in Cassano \& Brunetti 2005, C06 and C09) to derive the expected distribution of radio halos in the $P(120)-L_X$ diagram. 
The cosmological evolution of the magnetic field in this model is accounted for by scaling the field with the cluster mass, as suggested by cosmological MHD simulations (\eg Dolag et al. 2002; see also Sect.2). On the other hand, our calculations do not follow self-consistently turbulence and the amplification of magnetic fields due to this turbulence. The main reason for that is that in the turbulence acceleration scenario radio halos are generated and disappear due to the acceleration and cooling of the emitting particles, and such processes are much faster ($\sim0.1-0.2$ Gyr) than the slow decay of the magnetic field in the ICM, $\sim$ few Gyr (\eg Brunetti et al. 2009). Moreover, although particle-acceleration is demonstrated to be connected with cluster mergers (\eg Buote 2001; Govoni et al. 2004; Venturi et al. 2008), several mechanisms/sources of magnetic field, other than mergers-induced amplification, may significantly contribute to the magnetic field in the ICM. As a matter of fact stronger values of magnetic fields are measured in cooling core clusters, that are not merging systems (\eg Carilli \& Taylor 2002; Govoni 2006).
Although it is clear that a self-consistent treatment of turbulence, particle acceleration and magnetic field evolution is mandatory and deserves future theoretical efforts, in this Section we show that the results presented in this paper are not substantially affected by the evolution of the magnetic field when clusters become more relaxed after a merging phase.

In order to investigate the effect of an evolving magnetic field on our results, we consider the simulations of the generation and decay of dynamo-active turbulence developed by Subramanian, Shukurov \& Haugen (2006, hereafter SSH06). They studied the decay phase of an induced turbulent flow and magnetic field, that follows a saturation phase after the driving is switched off. They found that after the exponential growth of the magnetic field a saturation phase follows, then the turbulent energy and the magnetic energy decay in such a way that after an eddy turnover timescale the turbulent energy density is $\approx 1/2$ of its saturation value, while the magnetic field strength is still at $\approx90$\% of the saturation value (see Fig.~2 in SSH06).
By considering random motion with a typical initial speed $v_0=500$ km/s and scale $l_0\simeq 300$ kpc, which are appropriate for cluster-merger driven turbulence, the eddy turnover timescale is $t_0\simeq 0.6$ Gyr. Since we are interested in the evolution of radio halos on timescales of $\sim$ Gyr, and the radiative lifetime of the emitting particle is of $\sim 0.1-0.3$ Gyr, we can estimate the variation of the frequency $\nu_s$ as a function of time by adopting the classical formula $\nu_s\propto (B\chi^2)/(B^2+B_{cmb}^2)^2$ (see also Sect.~2). The ratio between $\nu_s$ in the case of time-dependent magnetic field and in the case of constant magnetic field is reported in Fig.~\ref{Fig.evolv_B} (left panel), where $t_0= 1.1$ Gyr (since here we have considered a saturation phase of $\Delta t=0.5$ Gyr before the driving is switched off).

\noindent 
Substantial differences are found only for $t>2$ Gyr when, however, the halo is expected to have already disappeared (being $\nu_s\propto \epsilon_t^4$, with $\epsilon_t$ being the turbulence energy density). The ratio between the steepening frequencies (Fig.~\ref{Fig.evolv_B}, left panel) at $t\sim t_0$ gives an estimate of the error we can make by neglecting the time-evolution of the magnetic field, that is $\leq 30\%$.

In the central panel of Fig.~\ref{Fig.evolv_B} we also report the time-evolution of the synchrotron spectra at different time ($t=0.9$, $1.1$, $1.5$, $1.8$ and $2$ Gyr, see figure caption) by assuming a constant magnetic field (as in the adopted model, solid lines) and an evolving magnetic field (as in SSH06, dashed lines). For $t\leq 1.1$ Gyr the difference in the monochromatic radio luminosities is less then 10\%. In addition, the right panel of Fig.~\ref{Fig.evolv_B} shows the evolution of the spectral index between 1.4 GHz and 330 MHz of these synchrotron spectra. Up to 1.2-1.3 Gyr, the difference in the spectral indeces is very small, less then 10\%.

\section{Conclusions}

The observed correlations between the halo radio power (at 1.4 GHz) and the cluster X-ray luminosity, mass and temperature, and the observed connection between radio halos and cluster mergers, suggest a link between the gravitational process of cluster formation and the generation of radio halos. Radio halos are likely generated during cluster-cluster mergers where a fraction of the gravitational energy dissipated is channelled into the acceleration of relativistic particles. A crucial expectation of the {\it turbulent re-acceleration} scenario, put forward to explain radio halos, is that the synchrotron spectrum of halos is characterized by a cut-off at frequency $\nu>\nu_s$ with $\nu_s$ determined by the efficiency of the acceleration process. 
The presence of this cut-off causes a bias, so that present radio observations at $\sim$ GHz frequencies are expected to detect only the most efficient radio phenomena in clusters, leaving unexplored a large population of radio halos characterized by spectral cut-off at lower frequencies (\eg C06, Brunetti et al. 2008; C09). Future low frequency radiotelescopes as LOFAR and LWA are expected to unveil the populations of ultra steep spectrum radio halos, with $\nu_s<1$ GHz, in clusters, providing to test the idea of {\it turbulent re-acceleration}.  

\begin{figure}
\centerline{
%\begin{center}
\includegraphics[width=0.36\textwidth]{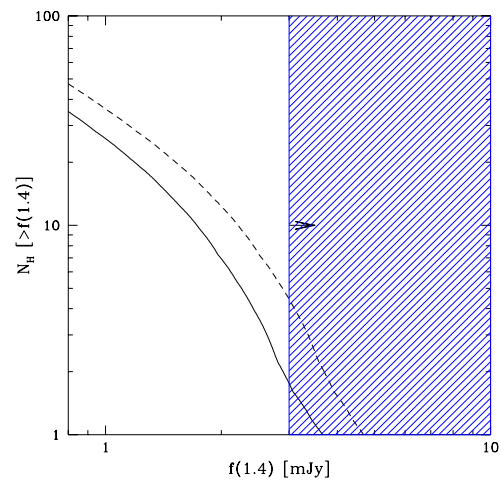}}
\caption[]{Integrated number counts at 1.4 GHz of halos with $600\leq\nu_s<1400$ MHz that are expected to be detected by LOFAR at 120 MHz assuming $\xi\,F\simeq0.25$ mJy/beam. Calculations are derived by combining {\it i)} at $z<0.3$, the sky coverage and X-ray flux limit of eBCS and REFLEX clusters, and {\it ii)} at $z=0.3-0.6$, the sky coverage and X-ray flux limit of MACS clusters. Dashed and solid lines account for the uncertainty due to the finite frequency range of $\nu_s$ ($600-1400$ MHz) assumed in the calculations. The dashed blue region shows the typical minimum flux of halos detectable with present radio facilities (\eg the VLA considering C+D configurations) at 1.4 GHz (approximatively $f(1.4)>3-4$ mJy).}
\label{Fig.NH_z0p6_1p4GHz}
%\end{center}
\end{figure}

One may wonder whether halos with $\nu_s<1$ GHz could be detected by present radiotelescopes at 1.4 GHz. To address this point in Fig.~\ref{Fig.NH_z0p6_1p4GHz} we report the flux distribution at 1.4 GHz of halos with $600\leq\nu_s<1400$ MHz (that in our model have $\alpha\approx 1.7$ between 120 MHz and 1400 MHz) that are expected to be detected by LOFAR at 120 MHz assuming $\xi\,F\simeq 0.25$ mJy/beam. Calculations are derived by assuming: {\it i)} at $z<0.3$ the X-ray flux limit and sky coverage of the extended {\it ROSAT} Brightest Cluster Sample (eBCS, Ebeling et al. 1998, 2000) and of the ROSAT-ESO Flux Limited X-ray Galaxy Cluster Survey (REFLEX, B\"oringher et al. 2004), and {\it ii)} at $z=0.3-0.6$ the X-ray flux limit and sky coverage of the Massive Cluster Survey (MACS, Ebeling et al. 2001).
Calculations show that potentially very deep pointed observations at 1.4 GHz of all these clusters may lead to the detection of a few of these ultra steep spectrum halos (those with $600\leq\nu_s<1400$ MHz); the ultra steep spectrum halo detected in the cluster Abell 521 (Brunetti et al. 2008; Dallacasa et al. 2009) belong to this class of halos, although it is among the flatter spectrum objects in this class, with $\nu_s\approx 1200$ MHz.

In this paper, we discuss the consequence of this new population of radio halos on the slope of the radio--X-ray luminosity correlation at low frequency. According to homogeneous models, ultra-steep spectrum halos are expected to be less luminous than halos with larger $\nu_s$ associated with clusters of the same mass. Also, radio halos with smaller $\nu_s$ should be statistically generated in clusters with smaller mass (and $L_X$). The combination of these two expectations implies that the radio-- X-ray luminosity correlation should be broader and steeper at lower frequencies.

Based on this model, we perform Monte Carlo simulations of the distribution of radio halos in the $P(120)-L_X$ plane. 
We find that halos distribute in the $P(120)-L_X$ plane according to a correlation which is steeper ($\Delta \alpha\approx0.4$) and broader than that observed at 1.4 GHz, with ultra-steep spectrum halos 
broadening the scatter in the region of low luminosity. We find that the number of ultra-steep spectrum halos increases with increasing the survey sensitivity and this further steepens the correlation.
The forthcoming LOFAR surveys should constrain the expected steepening of the correlation and test our expectations.

Although a self-consistent treatment of turbulence acceleration and amplification of the magnetic field in clusters is mandatory and deserve future efforts, we show that the main ingredients in the adopted scenario are two ``fast'' processes: particle acceleration and particle cooling that follow the decay of turbulence. Being a ``slow'' process, we show that the possible decay of the field with turbulence is not expected to affect the modeling of halo statistics significantly.

\begin{acknowledgements}
This work is partially supported by grants PRIN-INAF 2007, PRIN-INAF 2008 and ASI-INAF I/088/06/0. 
R.C. thanks the anonymous referees for comments and suggestions, and G. Brunetti, M. Br\"uggen, H.J.A. R\"ottgering and T. Venturi for useful comments. 
\end{acknowledgements}

\end{document}